\documentstyle[12pt,pra,aps]{revtex}

\begin{document}

\draft

\begin{flushright}
   CU-TP-676\\
   DFPD 95/NP/20
\end{flushright}
\vspace{1cm}
\begin{center}
{\Large \bf Strangeness Enhancement in $p+A$ and $S+A$\\
             Interactions at SPS Energies}
\footnotetext{*This work was supported by the Director,
Office of Energy Research,Division of Nuclear Physics of the Office
of High Energy and Nuclear Physics of the U.S.Department of Energy
under Contract No.DE-FG02-93ER40764}\\

   {\bf V.Topor Pop}$^1$\footnotetext{1. Permanent address
      Institute of Atomic Physics,
     Institute for Gravity and Space Sciences
     P.O.Box MG-6,Bucharest,{\bf Romania} E-MAIL TOPOR@ROIFA.BITNET;\\
     TOPOR@PADOVA.INFN.IT}\\
   {\normalsize Department of Physics,Columbia University,{\bf New York},
    N.Y.10027}\\
   {\normalsize and} \\
    {\normalsize Dipartamento di Fisica "G.Galilei",Via Marzolo 8-35131,
    Padova {\bf Italy} }\\
   {\bf M.Gyulassy}\\
   {\normalsize Department of Physics,Columbia University,{\bf New York},
   N.Y.10027}\\
   {\bf X.N.Wang}\\
   {\normalsize Nuclear Science Division,Lawrence Berkeley Laboratory,\\
   University of California,{\bf Berkeley},CA 94720}\\
   {\bf A.Andrighetto,M.Morando,F.Pellegrini,R.A.Ricci,G.Segato}\\
   {\normalsize Dipartamento di Fisica "G.Galilei",Via Marzolo 8 -35131,
   Padova {\bf Italy}}\\
   {\normalsize and }\\
   {\normalsize  INFN Sezione di Padova,{\bf Italy}}\\
   \end{center}
   \vskip 1cm
   \begin{center}
   {\it Work to be submitted to Phys.Rev.{\bf C}}
   \end{center}

   \newpage

\title{{\bf Strangeness Enhancement in $p+A$ and $S+A$\\
             Interactions at SPS Energies}$^*$\footnote{
This work was supported in part by the Director, Office of Energy
Research,Division of Nuclear Physics of the Office
of High Energy and Nuclear Physics of the U.S Department of
Energy under Contract No.DE-FG02-93ER40764.}}
\author{{\bf V.Topor Pop}\thanks{On leave from absence from Institute
 for Space Sciences,P.O.Box MG - 6,Bucharest,
{\bf Romania} E-MAIL TOPOR@ROIFA.BITNET(EARN
);TOPOR@PADOVA.INFN.IT}}
\address{\normalsize Department of Physics,Columbia University,{\bf New York},
N.Y.10027
\\
\normalsize and \\
\normalsize Dipartamento di Fisica "G.Galilei",Via Marzolo 8-35131,
Padova {\bf Italy}}
\author{\bf M.Gyulassy}
\address{\normalsize Department of Physics,Columbia University,{\bf New York},
N.Y.10027}
\author{\bf X.N.Wang}
\address{\normalsize Nuclear Science Division,Lawrence Berkeley Laboratory,\\
University of California,{\bf Berkeley},CA 94720}
\author{{\bf A.Andrighetto,M.Morando,F.Pellegrini,R.A.Ricci,G.Segato}}
\address{\normalsize Dipartamento di Fisica "G.Galilei",Via Marzolo 8 -35131,
Padova {\bf Italy}\\
\normalsize and \\
\normalsize  INFN Sezione di Padova,
{\bf Italy}}

\date{17 February 1995}

\maketitle

\begin{abstract}
The systematics of strangeness enhancement
 is calculated using the HIJING and VENUS models
and compared to recent data on  $\,pp\,$, $\,pA\,$
 and $\,AA\,$ collisions at CERN/SPS
 energies ($200A\,\, GeV\,$). The HIJING model is used
to  perform a {\em linear} extrapolation from
$pp$ to $AA$. VENUS is used to estimate the effects
of final state cascading and  possible non-conventional
production mechanisms. This comparison shows that
 the large enhancement of
strangeness  observed in $S+Au$ collisions,
interpreted previously as possible evidence for
quark-gluon plasma formation, has
its origins  in   non-equilibrium dynamics
of few nucleon systems.
  A factor of two  enhancement of $\Lambda^{0}$ at mid-rapidity
 is indicated by recent $pS$ data, where
 on the average {\em one} projectile nucleon
interacts with only {\em two} target
nucleons. There appears to be  another factor of
two enhancement in the light ion
reaction  $SS$ relative to $pS$,  when
on the average only two projectile
nucleons interact with two target ones.

\end{abstract}

\pacs{25.75+r ; 24.85.+r ; 24.85.+p ; 25.40.-h ; 25.40.Ve ; 24.10Jv}

\section{Introduction}

 The search for  new states of dense nuclear matter is one of the
most active areas of research in
nuclear physics \cite{pro6},\cite{pro5} .
Enhanced strangeness production  in ultra-relativistic
heavy ion collisions  was suggested long ago \cite{raf1}
as a signal for
quark-gluon plasma formation\cite{quer1},\cite{quer2},
and has been observed at both the AGS and SPS.
There is  extensive  data from both
 the SPS at CERN \cite{1na35}-\cite{mur94}
 and the AGS at BNL \cite{3e802}-\cite{bare94}
on strangeness yields from reactions ranging from elementary $p+p$
to $p+A_T$ and $A_B+A_T$ for targets ranging up to $A_T\approx 200$
and beams up to $A_B=30$. Detailed rapidity and transverse
momentum spectra of  ($K^{+}$,$K^{-}$, $K^{0}_{s}$,$\Lambda$
  ,$\bar{\Lambda}$)
are available and spectra of $\,\Xi^{-}\,$ and even $\,\Omega^{-}\,$
are becoming available.
In all  cases their yield relative to pions
or negative hadrons  are larger in nucleus-nucleus
than expected from geometrically  scaled   proton-proton collisions.
New experiments with truly heavy ion projectiles are in progress
with  Au beams at BNL\cite{dieb93},\cite{nam94}
and with Pb beams at CERN ($Pb(170A\, GeV)+Pb$) \cite{quer2}
and will soon extend considerably the data base.

These and other data on nuclear reactions
 have stimulated the development of  many hadronic transport models
to address the problem of multi-particle production
 in nuclear collisions.
These include Dual Partons Models(DPM) \cite{dpm94}-\cite{ranft4},
Quark Gluon String Models(QGSM)\cite{kai1}-\cite{ame2},
 VENUS  \cite{wer7}, FRITIOF \cite{ander}-\cite{nils},
  ATTILA \cite{Gyu1}, HIJING\cite{wang0}-\cite{wang4}
 RQMD models \cite{aich1}-\cite{sor94}, Parton String Model
 (PSM)\cite{ame94},HIJET\cite{fol94}, Parton Cascade
 Model(PCM) \cite{gei1}-\cite{gei3}.
 An excellent review and detailed comparison of the models
is given  by Werner in Ref.\cite{wer7}.

  At present no conventional explanation of the large
enhancement  of hyperons or antihyperons
  has been found.
 The  Pomeron exchange picture has motivated
the development of many of the above models with
  the Pomeron modeled in terms of
 colored strings. However, the  string picture itself
 suggests the possibility of new dynamical mechanisms
 ranging from string fusion to color rope formation.
Some of the above transport models like RQMD\cite{sor94}
and VENUS\cite{wer7}
include such non-conventional mechanisms as default options.
  These proposed novel {\em non-equilibrium}
 dynamical mechanisms were shown to be
able to reproduce many features of the observed strangeness enhancement
\cite{braun4}-\cite{merino2}, \cite{sor93a},\cite{sor94a}, \cite{wer7}.
On the other hand, there have been many attempts (see, e.g.,  review by Heinz
in \cite{pro6} p. 205c and references therein)
to attribute the strangeness enhancement
to the formation of an  equilibrated fireball containing
a quark-gluon plasma state \cite{pro6},\cite{rafel}.

Therefore it appears that either non-conventional
multi-particle mechanisms or the
existence of a new form of matter seems to be indicated by the
observed strangeness enhancement.
Either case is of basic  interest. The goal of the present
study is to clarify which
of these alternatives is more compelling.
We use the HIJING model\cite{wang0}-\cite{wang4}
to perform a {\em linear} extrapolation
of strangeness production dynamics from  $pp$ to $AA$
taking into account essential  nuclear geometry
and kinematical constraints. At higher collider energies
it includes  pQCD semi-hard processes, but in the SPS range
it reduces essentially to a hybrid version of the FRITIOF and DPM
models.
We use the VENUS model\cite{wer7} to estimate possible effects of
final state cascading and new  mechanisms
of strangeness production in few nucleon processes.
The non-conventional mechanism in VENUS4.13 is the
occurrence  of  ``double strings'' which may form when
one projectile nucleon interacts with two or more target nucleons.
A double string is defined as a color singlet baryon configuration
consisting of one projectile
quark connected to two different valence quarks in the target
via a three gluon vertex. In earlier versions of the model
the parameterization of the vertex kinematics led to anomalously
large baryon stopping power. In the present version,
the double string phenomenology is constrained to reproduce the
$pA\rightarrow pX$ data. However,
the new feature, see eq. 15.52 in ref. \cite{wer7},
is the assumption that the probability
for hyperon production in the fragmentation
 regions is enhanced
  by a factor of two relative to the single string rates.
This enhanced strangeness production  mechanism
due to  double strings is  similar to that postulated in
the color rope model\cite{biro} and incorporated into the RQMD model.
The hyperon enhancement in VENUS  is however more confined to the
fragmentation regions.

Both HIJING and VENUS have  been compared
to  a wide variety of data in
$\,pp\,$,$\,pA\,$ and $\,AA\,$ collisions  \cite{wang2},\cite{wang3},
\cite{wer7}.
However, no systematic study
 of  strangeness production at SPS-CERN energies
were performed up to now. In addition, there have been
substantial changes in the final published data \cite{na3594}
relative to earlier comparisons to preliminary data \cite{1na35},
\cite{2na35}.
 In  this paper, we calculate the  rapidity and transverse momentum spectra
of strange particles
for  $\,pp\,$, minimum bias collisions of
$\,pS,pAg,pAu\,$ and central collisions of
$S+S,Ag,Au,W\,\,$ at the energy of $\,200\,\,AGeV$
and $Pb+Pb$ at the energy of $\,170\,\,AGeV$.
We focus special emphasis on the comparison
with the data on $pp$, $pS$, and $SS$
 from Alber et al. \cite{na3594}.
That comparison reveals that
 much of the enhancement of strangeness
in heavy ion collisions can be traced back to
the  enhancement of strangeness  in the lightest nontrivial
ion collisions, $p+S$. Our main  conclusion based on these data
is that the enhancement of strangeness observed  in $S+Au$
is therefore most likely due to new non-equilibrium multi-particle
production mechanisms in processes involving {\em few} nucleon systems.

This paper is organized as follows:
A brief description of the HIJING Monte Carlo model and theoretical
background are given in Section II. For a detailed discussion of the
VENUS model,
we refer to the review in \cite{wer7}.
In Section III, detailed numerical results with
 HIJING and VENUS for $\,pp\,$,$\,pA\,$ and $\,AA\,$
reactions  at CERN-SPS energies  ($\,\sqrt{s}\simeq 20A \,\,GeV\,$) for
strangeness production are compared to experimental data and other
 model predictions .
 Section IV concludes with a summary and discussion of results.

\section{Outline of the HIJING Model}

A detailed discussion of the HIJING Monte Carlo model was reported
in references\cite{wang0}-\cite{wang4}.
The formulation of HIJING was guided by the LUND-FRITIOF and Dual
Parton Model(DPM)  phenomenology for soft nucleus-nucleus reactions at
intermediate energies ($\sqrt{s}<20\,\, GeV$) and implementation
of perturbative QCD(PQCD) processes in the PHYTHIA model\cite{sjos94}
 for hadronic interactions.
We give in this section a brief review of the aspect of the model relevant to
hadronic interaction:

\begin{enumerate}
\item Exact diffuse nuclear geometry is used to calculate the impact
parameter dependence of the number of inelastic processes
\cite{Gyu1}.
\item Soft beam jets are modeled by quark-diquark strings with gluon
kinks along the lines of the DPM and FRITIOF models. Multiple low
$\,p_{T}\,$ exchanges among the end point constituents are included.
\item  The model includes multiple mini-jet production with initial
and final state radiation along the lines of the PYTHIA model
 and with cross sections calculated
within the eikonal formalism.
\item Hadronization is performed via the JETSET7.2 algorithm \cite{sjos94}
that summarizes data on $e^+e^-$.
\item HIJING does not incorporate any  mechanism for final state
interactions among low $\,p_{T}\,$ produced particles
nor does it have color rope formation.

\end{enumerate}

The rate of multiple mini-jet production in HIJING is constrained by
the cross sections in nucleon-nucleon collision. Within an eikonal
formalism \cite{hwa1} the total elastic cross sections
$\sigma_{el}$,total inelastic cross sections $\sigma_{in}$ and
total cross sections $\sigma_{tot}$ can be expressed as:
\begin{equation}
\sigma_{el}=\pi\int_{0}^{\infty}\:db^{2}(1-\exp(-\chi(b,s)))^{2}
\label{e1}
\end{equation}
\begin{equation}
\sigma_{in}=\pi\int_{0}^{\infty}\:db^{2}(1-\exp(-2\,\chi(b,s)))
\label{e2}
\end{equation}
\begin{equation}
\sigma_{tot}=2\,\pi\int_{0}^{\infty}\:db^{2}(1-\exp(-\chi(b,s)))
\label{e3}
\end{equation}
Strong interactions involved in hadronic collisions can be
generally divided into two categories depending on the scale of
momentum transfer $q^{2}$ of the processes.
If $q^{2}< \Lambda_{QCD}^{2}$ the collisions are
nonperturbative and  considered {\it soft}\,\, and modeled
by beam jet fragmentation via the string model. If
$q^{2}\gg \Lambda_{QCD}^{2}$ the subprocesses on the parton
level are considered {\it hard} and  calculated via pQCD
\cite{wang3}.

In the limit that the real part of the scattering amplitude is
small and the eikonal function $\chi(b,s)$ is real, the factor
\begin{equation}
g(b,s)=1-\exp(-2\,\chi(b,s))
\label{e4}
\end{equation}
can be interpreted in terms of semi-classical probabilistic model  as
{\sl the probability for an inelastic event of nucleon--nucleon
collisions at impact parameter $b$}  which may be caused by hard,
semi-hard or soft parton interactions.

To calculate the probability of multiple mini-jet, the main dynamical
assumption is that they are independent. This holds as long as
their average number is not too large as is the case below
 LHC energies \cite{wang3}.
When shadowing can be neglected,the probability of no jets and
 $j$ independent jet production in an inelastic event at impact
parameter $b$,can be written as :
\begin{equation}
g_{0}(b,s)=(1-\exp(-2\,\chi_{s}(b,s)))\exp(-2\,\chi_{h}(b,s))
\label{e5}
\end{equation}
\begin{equation}
g_{j}(b,s)=\frac{\left[2\,\chi_{h}(b,s)\right]^{j}}{j!}\cdot
            exp(-2\,\chi_{h}(b,s))\,\,\,\,\,\,  j \geq 1
\label{e6}
\end{equation}
where $\chi_{s}(b,s)$ --is the eikonal function for soft interaction,
$2\,\chi_{h}(b,s)$ --is the average number of hard parton interactions
at a given impact parameter,
$\exp(-2\,\chi_{s}(b,s))$ --is the probability for no soft interaction.
Summing eqs.(5) and (6) over all values of $j$ leads to :
\begin{equation}
\sum_{j=0}^{\infty}\,\,g_{j}(b,s)=1-\exp(-2\,\chi_{s}(b,s)-2\,
\chi_{h}(b,s))
\label{e7}
\end{equation}
Comparing with eq.(\ref{e4}) one has :
\begin{equation}
\chi(b,s)=\chi_{s}(b,s)+\chi_{h}(b,s)
\label{e8}
\end{equation}

Assuming that the parton distribution function is factorizable
in longitudinal and transverse directions and that the shadowing can be
neglected the average number of hard interaction $2\chi_{h}(b,s)$
at the impact parameter $b$ is given by :
\begin{equation}
\chi_{h}(b,s)=\frac{1}{2}\,\,\sigma_{jet}(s)\,T_{N}(b,s)
\label{e9}
\end{equation}
where $T_{N}(b,s)$ is the effective partonic overlap function of the
nucleons at impact parameter $b$.
\begin{equation}
T_{N}(b,s)=\int d^{2}b'\rho(b')\rho(\left|b-b'\right|)
\label{10}
\end{equation}
with normalization
$\int\,\,d^{2}b\,T_{N}(b,s)=1$
and $ \sigma_{jet}\,\,$ is the pQCD cross section
of parton interaction
or jet production \cite{wang2},\cite{wang3}.
Note that $\,\,\xi=b/b_{0}(s)$, where $b_{0}(s)$ provides a
measure of the geometrical size of the nucleon
$\pi b_{0}^{2}(s)=\sigma_{s}(s)/2$
assuming the same geometrical distribution for both
soft and hard overlap functions
\begin{equation}
\chi_{s}(\xi,s)\equiv \frac{\sigma_{s}}{2\sigma_{0}}\chi_{0}(\xi)
\label{e11}
\end{equation}
\begin{equation}
\chi_{h}(\xi,s)\equiv \frac{\sigma_{jet}}{2\sigma_{0}(s)}\chi_{0}(\xi)
\label{e12}
\end{equation}
\begin{equation}
\chi(\xi,s)\equiv \frac{1}{2\sigma_{0}}\left [{\sigma_{s}(s)
+\sigma_{jet}(s)}\right ]\chi_{0}(\xi)
\label{e13}
\end{equation}
We note that $\,\chi(\xi,s)\,$ is a function not only of $\,\xi\,$
 but also of
$\,\sqrt{s} \,$ because of the $\,\sqrt{s}\,$ dependence on
the jet cross
section $\,\sigma_{jet}(s)\,$.Geometrical scaling implies on the other
hand that $\,\chi_{s}(\xi,s)=\chi_{0}(\xi)\,$ is only a function
 of $\,\xi\,$.
Therefore, geometrical scaling is broken at high energies by the
introduction of $\,\sigma_{jet}(s)\,$ of jet production.

The cross sections of nucleon - nucleon collisions
can in this case be expressed as:
\begin{equation}
\sigma_{el}=\sigma_{0}(s)\int_{0}^{\infty}d\,\xi^{2}\left ( 1-exp(-\chi
(\xi,s)\right )^{2}
\label{e14}
\end{equation}
\begin{equation}
\sigma_{in}=\sigma_{0}(s)\int_{0}^{\infty}d\,\xi^{2}\left ( 1-exp(-2\,
\chi(\xi,s)\right))
\label{e15}
\end{equation}
\begin{equation}
\sigma_{tot}=2\,\sigma_{0}(s)\int_{0}^{\infty}d\,\xi^{2}\left (1-exp
(-\chi(\xi,s)\right))
\label{e16}
\end{equation}
The calculation of these cross sections requires specifying
$\sigma_{s}(s)$ with a corresponding value of cut - off momenta
$p_{0}\approx 2$ GeV/c \cite{wang4}.

In the energy range $10\,\, GeV<\sqrt{s}<70\,\, GeV$, where only soft
parton interactions are important, the soft  cross section
$\sigma_{s}(s)$ is fixed by the data on total cross sections
$\sigma_{tot}(s)$ directly. In and  above the $Sp\bar{p}S$ energy range
$\sqrt{s}\geq 200\,\, GeV$,  a fixed $\sigma_{s}(s)= 57$ mb
 and a mini-jet cutoff scale $p_{0}=2 \,\,GeV/c$,  leads to
observed energy dependence of  the
cross sections and inclusive distributions. Between the two regions
$70\,\, GeV < \sqrt{s}<200\,\, GeV$,  a smooth extrapolation
for $\,\sigma_{s}(s)\,$ is used.

In HIJING, a nucleus-nucleus collisions is decomposed into
a sequence of binary
collisions involving in general excited or wounded nucleons.
Wounded nucleon are assumed to be $\,q-qq\,$ string like
configurations that decay on a slow time scale compared to the
collision time of the nuclei. In the FRITIOF scheme wounded nucleon
interactions follow the same excitation law as the original hadrons.
In the DPM scheme subsequent collisions essentially differ from
the first since they are assumed to involve sea partons instead
of valence ones. The HIJING model adopts a hybrid scheme, iterating
string-string collisions as in FRITIOF but utilizing DPM like
distributions. In the SPS range the HIJING results for nuclear collisions
are  very similar to those of  FRITIOF.
However, HIJING provides an interpolation model
 between the  nonperturbative beam jet fragmentation physics
at intermediate CERN-SPS energies and perturbative QCD mini-jet physics
at the highest collider energies ($\,RHIC\,,\,LHC\,$).

\section{NUMERICAL RESULTS}
\subsection{STRANGENESS IN PROTON - PROTON INTERACTION}

We used  the program HIJING with default parameters:
 IHPR2(11)=1 gives the baryon production
model with diquark-antidiquark pair production allowed, initial
diquark treated as unit; IHPR2(12)=1, decay of particle such as
$\,\pi^{0}\,$,$\,K_{s}^{0}\,$,$\,\Lambda\,$,$\,\Sigma\,$,
$\,\Xi\,$, $\,\Omega\,$ are allowed\,\,;\,\,IHPR2(17)=1
 - Gaussian distribution of transverse
momentum of the sea quarks ;IHPR2(8)=0 - jet production turned
off for theoretical predictions denoted by HIJING model,
 and IHPR2(8)=10-the maximum
number of jet production per nucleon-nucleon interaction
for  for theoretical predictions denoted by
$\,HIJING^{(j)}\,$ for comparison.

In Table I the calculated average multiplicities of particle at
$E_{lab}=200\,\, GeV\,$ in proton-proton($pp$) interaction
are compared to data.
The theoretical values $\,HIJING\,$ and  $\,HIJING^{(j)}\,$
are obtained for
$\,\,10^{5}\,\,$ generated events and in a full phase space.
The values $\,\,HIJING^{(j)}\,\,$ include the
very small possibility of mini jet production at these low SPS energies.
The experimental data are taken from Gazdzicki and Hansen
\cite{11na35}.

The small kaon to pion ratio is due to the suppressed
strangeness production basic to string fragmentation. Positive pions and
kaons are  more abundant than the negative ones due to
charge conservation. We note that the {\em integrated}
multiplicities for neutral
strange particle $\,\,<\Lambda>,<\bar{\Lambda}>,<K_{s}^{0}>\,\,$
are reproduced at the level of three standard  deviations for
 $\,pp\,$ interactions at  $\,200\, GeV\,$. However the values for
$\,<\bar{p}>\,$  and $\,<\bar{\Lambda}>\,$ are significantly
 over predicted  by the model. This is important since
as we shall see the $\bar{\Lambda}$ in $S+S$ is significantly underestimated
by HIJING.

For completeness
we include a comparison of hadron yields
at collider energies$\,\,\sqrt{s}=546\,\, GeV$($\,Sp\bar{p}S\,$-energies),
for $\,\,\bar{p}p\,\,$ interactions, where mini-jet production plays a much
more important role.
{}From different collider experiments
Alner et al.(UA5 Collaboration))\cite{1ua5}  attempted to piece
 together a picture of the composition of a typical soft event at the
$\,Sp\bar{p}S\,$  \cite{ward}. The measurements were made in various
different kinematic regions and have been extrapolated in the full
transverse momenta($\,p_{T}\,$)  and rapidity range for comparison
as described in reference \cite{1ua5}.
The experimental data are compared to theoretical values obtained with
$HIJING^{(j)}$  in Table II.
 It was stressed  by  Ward \cite{ward} that
 the data show a substantial excess of photons compared to the
 mean $\,\,\pi^{+}+\pi^{-}\,\,$. It was suggested as a possible
 explanation of such enhancement a  gluon
 Cerenkov radiation emission in hadronic collision
\cite{drem}. Our calculations rules out such hypothesis.
 Taking into account decay from resonances and direct gamma production,  good
agreement is found within the experimental errors.
 The experimental ratio $\,\frac{K^{+}}{\pi^{+}} =0.095 \pm 0.009\,$
 is also reproduced by$\,\, HIJING^{(j)}\,\,$ model (0.099).
 We  note that a detailed
 study of the ratios of invariant cross sections of kaons to that
 of pions as a function of transverse momenta in the central region
 was presented in \cite{wang3}.

In the following plots
the kinematic  variable used to  describe single particle properties are
 the transverse momentum $\,p_{T}\,$ and the rapidity $\,y\,$ defined
as usual as:
 \vskip 0.3cm
 \begin{equation}
 y=\frac{1}{2}ln \frac{E+p_{3}}{E-p_{3}}=ln\frac{E+p_{3}}{m_{T}}
 \label{e17}
 \end{equation}
 \vskip 0.3cm
 with $\,E,p_{3}\,$,and $\,m_{T}\,$ being energy,longitudinal momentum and
 transverse mass
$ m_{T}=\sqrt{m_{0}^{2}+p_{T}^{2}}$
 with $\,m_{0}\,$ being the particle rest mass.

In Fig.1a, 3a, 4a, and 6a,  we show rapidity  and
 transverse momentum distributions
 for $\,\,\Lambda\,\,$'s
 (Fig.1a,4a) and  $K_{s}^{0}$'s (Fig.3a,6a)
  produced in  $ pp$ scattering at$\,\, 200\,\,$ GeV.
  The theoretical histograms obtained with
 HIJING (solid) and VENUS-4.13 (dashed)
are compared with experimental
 data taken from Jaeger et al.\cite{jagl}.
The  HIJING spectra for $\,\Lambda\,$,$\,K_{s}^{0}\,$
are  close to the data at mid rapidity \cite{jagl},
although the dip in the $\,\,K_{s}^{0}\,\,$ yield at mid-rapidity
and the $\,\,\Lambda\,\,$ peak in the fragmentation regions
are not well reproduced (see
 also ref. \cite{wer7}). Unfortunately, more precise
 data are not  available in $pp$ interactions
and  those features could reflect experimental acceptance cuts.
Similarly no detailed $\bar{\Lambda}$ spectra are as yet available
in $pp$.

In comparison with VENUS (taking $10^4$ events)
we note that this version seems
to over-predict the $pp\rightarrow \Lambda^0$ rapidity density at mid-rapidity
by $50-100\%$ in Fig. 1a, even though the rapidity integrated
transverse momentum distribution in Fig 4a seems closer to the
data. The $K_s^0$ yields in Figs. 3a and 6a are similar to those of HIJING
with the dip structure in the data absent.

The very sparse  data base on  $pp$ strangeness production
at  SPS energies should be expanded in the future to improve
the test of dynamical models before they are applied to the more
complex nuclear collision case.
Without $\,\bar{\Lambda}\,$ spectra in $pp$, for example, the
need for the new dynamical mechanisms in that channel cannot
be confirmed.

\subsection{Multiplicities in $pA$ and $AA$ collisions}

 In this section, we compare  strange particle production
 in the HIJING and VENUS models to $pA$ and $AA$
data.
Again we limit the study to
 $\,\,\Lambda,\bar{\Lambda},K_{s}^{0}\,\,$
 to  compare with recent  data from
  Alber et al.\cite{na3594}.
 First we consider the average integrated multiplicities
 for negative hadrons $\,<h^{-}>\,$, negative pions $\,<\pi^{-}>\,$
 and neutral strange particles $\,<K_{s}^{0}>\,$,
$\,<\Lambda>\,$, $\,<\bar{\Lambda}>\,$ in $\,pp,pS,pAg,pAu\,$
(minimum bias collisions) and  $\,SS,SAg,SAu\,$
(central collisions) at  $\,200$ AGeV.
 The default parameters of HIJING were used without mini-jet production
(IHPR2(8)=0).
 The number of Monte Carlo generated events was $\,10^{5}\,$ for HIJING
and $\,10^4\,$ for VENUS for $pp$,$pA$ interactions and
$\, 5\cdot 10^{3}$
for $\,SS,$ and $\,10^{3}\,$ for $\,S+Ag,W,Au\,$
and $\,PbPb\,$ collisions.

 The mean multiplicities
 are compared  in Table III (for $pp$ and $pA$ interactions) and in
 Table IV (for $AA$ interactions) with
 experimental data from Alber et al.\cite{na3594}.
 Note that
while the HIJING model describes well the integrated neutral
 strange particle multiplicities (except  for $<\bar{\Lambda}>$)
in $pp$ and  $pA$ interactions, there is a large discrepancy
already   for the light ion $S+S$ reaction.

It is worthwhile to mention that theoretical calculations
 have been done  for  $\,\,pA\,\,$ 'minimum bias' collisions
 and the experimental data are
 for the events with charged particle multiplicity greater than
 five,which contain a significant fraction (about 90 \%) of the
 'minimum bias' events \cite{na3594}.

 In Table III and IV    the data are compared  also with
 other theoretical values obtained in some  models :
 VENUS (as computed here), RQMD \cite{na3594},
 QGSM \cite{ame2},\cite{na3594} and DPM models.
The theoretical values$\,DPM^{1}\,$ are
 from the  Mohring et al.\cite{mohr1}, version of DPM which include
 additionally  $\,\,(qq)-(\bar{q}\bar{q})\,\,$ production
 from the sea into the chain formation process and the values
  $\,DPM^{2}\,$ are from the Mohring et al. \cite{mohr2},
 version of DPM which include chain fusion , as a mechanism to
 explain the anomalous antihyperon production.

 Alber et al. \cite{na3594} have considered that  the total production
of strangeness should be treated in a model independent way using all
 available experimental information for ratio $\,\,E_{S}\,\,$
 expressed as :
\vskip 0.3cm
\begin{equation}
E_{S}=\frac{<\Lambda>+4<K_{s}^{0}>}{3<\pi^{-}>}
\label{e18}
\end{equation}

\vskip 0.3cm
  We have calculated  this ratio in HIJING approach for the above
 interactions and the corresponding numerical predictions
 are shown in Table V.
We note that there is much less discrepancy between HIJING and the data
for this particular ratio. We conclude from this that such
 ratio is insensitive to the underlying physics and
therefore  should NOT
be used for any further tests of models! This ratio
 hides very effectively the gross
deficiencies of the HIJING  model in $SS$ reactions
pointed out later in  the comparison to the
rapidity and transverse momentum distributions.
We include Table V  only to prove
 the futility of studying the systematics of such
ratios in the search for novel dynamics in nuclear collisions!

 \subsection{Single inclusive distributions for  neutral strange
 particles in $\,pA\,$ and $\,AA\,$ }

The main results of the present study.
are contained  in Figs. 1-6.
Figure 1 is our most important result revealing the systematics
of $\Lambda$ enhancement   from (a) $pp$ to (d) $SAu$.
In part (a) the pp data at mid rapidity are seen to be
well reproduced by HIJING.
 However,  the new minimum bias $pS$ data\cite{na3594} in Fig. 1b
clearly shows  a factor of $2-3$  discrepancy with respect to the linear
extrapolation
from $pp$ as performed by HIJING. The effect of double string
fragmentation and final state cascading,
as modeled  with VENUS is seen on the other hand to
account for the observed $\Lambda$ enhancement.
We note, however,  that in $pp$,
VENUS  over-predicts the $\Lambda$ yield at mid rapidity.
Some fraction of the aggreement in $pS$
with VENUS may be due to this effect.
The overprediction of midrapidity $\Lambda$'s in $pp$
by VENUS was shown in Fig. 10.20b of ref.\cite{wer7},
but was not emphasized there.
If both the $pp$ and $pS$ data on $\,\Lambda\,$
production are correct,then the most striking increase of hyperon
production therefore
occurs between  $pp$ to $pS$ reactions.

The strangeness enhancement in  minimum bias $p+S$
is striking because the  number of target nucleons
struck by the incident proton is on the average only two!
The step from single $p+p$ to triple $p+p+p$ reactions therefore
apparently leads a substantial enhancement of
midrapidity $\Lambda$'s which obviously cannot
 have anything to do with equilibrium physics.

In central $S+S$ reactions shown in Fig. 1c,
the discrepancy relative to HIJING grows by  another factor of two.
We note that the new data\cite{na3594}shown here  have increased
substatially relative to earlier data  \cite{1na35},\cite{2na35}
due to inclusion of lower transverse momentum regions
and $\Lambda$'s originating from the decay
of $\Sigma,\Xi$ in the analysis.
Including these decay channels, VENUS is seen to reproduce the
new data as well.
We note that with RQMD the excess $\Lambda$'s is also  reproduced with the
introduction of  rope formation (see Table IV).
For heavier targets, $\,Ag,Au\,$, in Fig. 1d,
the discrepancy relative to HIJING is in fact less dramatic
than in $\,S+S\,$.

In central $\,S+S\,$,  on the average each projectile nucleon
interacts with only two target one, but each target nucleon also interacts
with two projectile ones.
In effect, then  $\,S+S\,$ reactions probe  strangeness production in
four nucleon interactions  $p+p+p+p\rightarrow \Lambda + X$.
Such reactions appear to be approximately  four times
as efficient in producing   midrapidity $\Lambda$'s
as two nucleon interactions in part (a).
Our main  conclusion therefore is that
strangeness enhancement is a nonequilibrium dynamical effect
clearly revealed in the lightest ion interactions.

Further support for this conclusion is shown  in Figs. 4 a,b,c,
where  the  transverse momentum distributions are compared.
We see that there is an
enhancement of the $\Lambda$ transverse momentum
relative to  $pp$ in  $pS$. Comparing to VENUS we can interpret Fig. 4b
as evidence that the enhanced
transverse momentum of $\Lambda$ in $pS$ is due
cascading. The discrepency
in Fig. 4c between VENUS and the data in $SS$ may be due to the rapidity
cuts in the data, which we have not included in the calculated spectra.
In all cases the deficiency
of the linear extrapolation via the HIJING model
is clearly evident. For heavier targets,  $S+Ag,Au$,
the transverse momentum distribution predicted by VENUS
is close to the data.

The same general conclusion  emerges from the systematics of
$\bar{\Lambda}$ and $K_{s}^{0}$ production s in Figs 2,3 and in Figs 5,6
respectively. In Fig.2a, the agreement between HIJING and VENUS and
the data on the $p+S\rightarrow
\bar{\Lambda}$  must be viewed with caution
since as shown in Table III, both models overpredict the
integrated $\bar{\Lambda}$ multiplicity by a factor $2-3$.
Given the absence of more detailed rapidity and transverse momentum
distributions
for $\bar{\Lambda}$, it is not possible  to determine whether
the $pp$ and  $pA$ data are compatible.
However, at least the step form $pS$ to $SS$ in Fig 2b
indicates  a possible factor of two enhancement of $\bar{\Lambda}$
similar to  the comparison of Fig. 1b,1c for ${\Lambda}$.
As in the case of ${\Lambda}$ production, there appears to be no further
$\bar{\Lambda}$ enhancement from $SS$ to $SAu$.
As regards to the transverse momentum distributions in Fig. 5,
we note that as in Fig. 4 the $\bar{\Lambda}$ emerge with higher $p_\perp$
in $pS$ than in $pp$ in accord with the VENUS model.
We note that in Figs 5 c,d,  the norm theoretical curves
is obtained intergrating  over the full rapidity interval,
while the norm data are limited to a smaller domain as shown in Fig. 2 b,c.

In the case of  $K_{s}^{0}$ production in Figs 3, 6, the same general
trends are seen but in a  less dramatic form.

We conclude that the new data
indicate that the origin of strangeness enhancement
in heavy ion collisions may be traced back
to  non-conventional and necessarily
 non-equilibrium dynamical effects
that arise in collisions of three or more nucleons.
However, this conclusion is forced upon us by the sytematics
of the new light ion data on $p+S$ and $S+S$ reported in \cite{na3594}.
As shown in Figs 1b, 2a, and 3b, those systematics, especially in $pA$, differ
considerably from the trends of earlier NA5
data\cite{10na35} and preliminary NA36 data\cite{1ander94}-\cite{greiner95}.
Those data for {\em heavier}
target nuclei incidate substantially less enhancement of midrapidity
$\Lambda,\bar{\Lambda},K^0_s$ than do the NA35 data on  $p+S$.
Part of the difference between these data sets may be due to different
acceptance cuts and the inclusion or rejection
of fragments from decay of higher mass hyperons.
Obviously, the difference between these data sets
must be  resolved.
Until then, the NA36 data must be regarded as
an important  caveat on our conclusions.

For completeness we show also  in Figs 7 the linear extrapolations
of HIJING to $Pb+Pb$ at 170 AGeV for all positives (Fig 7a) and
all negatives charges (Fig 7b),
for $\Lambda$ (Fig 7c) and for $\bar{\Lambda}$ (Fig 7d).
 It will be interesting to compare these extrapolations with
upcoming data to test if the strangeness enhancement
increases from $SS$ to $PbPb$.

We include the two dimensional distributions
 in  Fig. 8 to emphasize that strangeness enhancement
analyses restricted to narrow rapidity and transverse momentum
cuts, especially with simplistic fireball models,
may completely miss the global non-equilibrium character of the data.

\section{Conclusions}

In this paper we performed a systematic analysis  of strange particle
production in $\,pp\,$,$\,pA\,$ and $\,AA\,$ collisions at
SPS CERN-energies using the HIJING and VENUS models.
The most surprising result is that  the breakdown of the linear
 extrapolation from $pp$ data to nucleus-nucleus
 in the strangeness channel already occurs
 in minimum bias $pS$ !
The apparent enhancement
of $\Lambda$, $\bar{\Lambda}$ and $K_{s}^0$
at midrapidities in $pS$ reactions by a factor of 2 indicates
 that the mechanism for strangeness enhancement
 in heavier ion collisions must be associated with
non-equilibrium dynamics involving multiparticle production
and not with equilibrium  quark-gluon fireball.
In minimum bias $pS$ one projectile nucleon interacts on the
average with only two target ones.
 The data\cite{na3594} on $pS$ therefore indicate the existence of new
dynamical mechanisms for strangeness production that becomes operative
in   $p+p+p$ collisions.
The new data\cite{na3594} on central $S+S$
show another  factor of 2 enhancement of strangeness production
relative to $pS$. This light ion reaction basically probes
multiparticle production in $p+p+p+p$.
The strangeness enhancement in heavier target systems
apparently saturates at the S+S level.
We also showed  that traditional analysis of strangeness
enhancement in terms of ratios of integrated multiplicities
is very ineffective  since those ratios hide well
 defects of  the detailed rapidity
and transverse momentum distributions predicted by models.

The agreement with VENUS and RQMD results suggests  color rope
formation as a possible mechanism.
However, to clarify the new physics
 much better quality
data on elementary $p+p$ as well as on other  light ion
$p+\alpha,C,S$ and $\alpha+\alpha,C,S$
reactions
will be needed. Especially, the discrepancy between NA35 and NA36
must be resolved.
Only then can strangeness enhancement systematics
 used meaningfully  in the search for signatures of  quark-gluon plasma
formation  in future experiments
with $Au+Au$ and $Pb+Pb$.

\section{Acknowledgements}

We are grateful to Klaus Werner for providing the source code of VENUS.
One of the authors (VTP) would like to  expresses  his
 gratitude to Professor C.Voci for kind invitation and
 acknowledge financial support from INFN-Sezione di Padova,
 Italy where  this work was initiated.
 VTP  is also indebted to Professor E.Quercigh for his kind hospitality
in CERN (May 1994), and for very useful discussions.
Finally, VTP greatfully acknowledges   partial financial support from
the  Romanian Soros Foundation for Open Society  Bucharest,Romania.

 \newpage
 \begin{figure}
 \caption{Rapidity distributions of $\,\Lambda^0\, $  produced in
 $pp$ interactions at $\,200\,\,GeV\,$ (Fig.1a).
 The data for $pp$ (black small circles)  are
 from Jaeger et al.\protect{\cite{jagl}}.
 Rapidity distributions of $\,\Lambda^0\, $  produced in
 minimum bias $\,pS\,$ (Fig.1b) and  central
 $\,SS\,$(Fig.1c),$\,SAg\,$(Fig.1d) and $\,SAu\,$ (Fig.1d)
 collisions at $\,200$ AGeV.
 HIJING and VENUS results are shown by solid and dashed histograms
(for $\,pp\,$,$\,pS\,$, $\,SS\,$, $\,SAu\,$).
The new NA35  data ($\,pS\,$,$\,SS\,$-full circles;
 $\,SAg\,$ - stars ; $\,SAu\,$-full triangle) are from Alber et al.
 \protect{\cite{na3594}}.
 The open circles show the distributions
 for $\,SS\,$ collisions reflected at $\,y_{lab}=3.0\,$.
In Fig 1b,  earlier  NA5 data on $p+Ar$  (open diamond)
from ref.\protect{\cite{10na35}}
and preliminary data on $p+Pb$  (open squares)
from NA36\protect{\cite{1ander94,greiner95}}
are  shown for comparison. In Fig1c NA36  data on
$S+S$(open squares) are also shown for comparison.}
 \label{fig1}
 \end{figure}

\begin{figure}
\caption{Rapidity distributions of  $\,\bar{\Lambda}\, $
 produced   in minimum bias  $\,pS\,$ (Fig.2a) and  central
 $\,SS\,$(Fig.2b),$\,SAg\,$(Fig.2c) and $\,SAu\,$ (Fig.2d)
 collisions at $\,200\,$ AGeV.
 Solid and dashed histograms are as in Fig.1 .
The NA35 data (full circles) are from Alber et al.
 \protect{\cite{na3594}}. The open circles show the distributions
 for $\,SS\,$ collisions reflected at $\,y_{lab}=3.0\,$.
In Fig. 2a, the open squares correspond to preliminary $p+Pb$ data from
NA36\protect{\cite{1ander94,greiner95}}.}
 \label{fig2}
 \end{figure}

\begin{figure}
 \caption{As in Fig.1 but for $\,K_{s}^{0}\,$ particles.\hspace{3.3in}}
\label{fig3}
 \end{figure}

\begin{figure}
 \caption{Transverse momentum  distributions of
 $\,\Lambda^0\, $ produced in $\,pp\,$ interactions
at $\,200\,$ AGeV (Fig.4a). The  data
 (black small circles)for $\,pp\,$ interactions are
from Jaeger et al.\protect{\cite{jagl}}.
 Transverse kinetic energy distributions of  $\,\Lambda^0\, $
produced in minimum bias $\,pS\,$ (Fig.4b) and  central
 $\,SS\,$(Fig.4c),$\,SAg\,$(Fig.4d) and $\,SAu\,$ (Fig.4d)
 collisions at $\,200\,$ AGeV.
HIJING and VENUS results are shown by solid and dashed histograms resp.
(for $\,pp\,$,$\,pS\,$, $\,SS\,$, $\,SAu\,$).
The NA35  data ($\,pS\,$,$\,SS\,$-full circles;
 $\,SAg\,$ - stars ; $\,SAu\,$-full triangle) are  from Alber et al.
 \protect{\cite{na3594}}.}
 \label{fig4}
 \end{figure}

\begin{figure}
 \caption{Transverse  kinetic energy distributions for
  $\,\bar{\Lambda}\,$ particles produced
in minimum bias $\,pS\,$ (Fig.5a) and  central
 $\,SS\,$(Fig.5b),$\,SAg\,$(Fig.5c) and $\,SAu\,$ (Fig.5d)
 collisions at $\,200\,\,GeV\,$ per nucleon.
 Solid  and dashed histograms are as in Fig.4.
The experimental data (full circles) are  from Alber et al.
 \protect{\cite{na3594}}.}
 \label{fig5}
 \end{figure}

\begin{figure}
 \caption{As in Fig.4, but for  $\,K_{s}^{0}\,$ particles .}
 \label{fig6}
 \end{figure}

\begin{figure}
\caption{Predicted rapidity distributions for central ($b=0-1$ fm)
$\,PbPb\,$ collisions at $\,170\,$  AGeV with the HIJING  model
 for all positive charges (Fig.7a),
all negative charges  (Fig.7b),   $\,\Lambda\,$ (Fig.7c)
 and  $\,\bar{\Lambda}\,$ (Fig.7d).
}
\label{fig7}
\end{figure}

\begin{figure}
\caption{Unnormalized rapidity $\,y\,$ and  transverse momentum
 $\,p_{T}\,$ distributions for $\,\Lambda\,$ (Fig.8a,b) and
$\,\bar{\Lambda}\,$ (Fig.8c,d)
for central $\,PbPb\,$ at 170 AGeV from HIJING.}
\label{fig8}
\end{figure}

\newpage

\begin{table}
\caption{Particle multiplicities for $\,pp\,$
  interaction at $\,\,200\,$  GeV are compared with
  data from Gazdzicki and Hansen\protect{\cite{11na35}}}
\label{Tab1}
\begin{tabular}{|c|c|c|c|}    \hline
{\bf pp} & {\bf Exp.data}  & {\bf HIJING} &${\bf HIJING^{(j)}}$ \\
\hline
 $<\pi^{-}>$  & $2.62\pm 0.06$ & $2.61$ & $2.65$ \\
 \hline
 $<\pi^{+}>$ & $3.22\pm 0.12$  & $3.18$ & $3.23$  \\
 \hline
 $<\pi^{0}>$  & $3.34 \pm 0.24$ & $3.27$ & $3.27$ \\
 \hline
 $<h^{-}>$  & $2.86 \pm 0.05$ & $2.99$ &$ 3.03$ \\
 \hline
 $<K^{+}>$  & $0.28 \pm 0.06$ & $0.32$ & $0.32$ \\
 \hline
 $<K^{-}>$  & $0.18 \pm 0.05$ & $0.24$& $0.25$  \\
 \hline
$<\Lambda + \Sigma^{0}>$ & $0.096 \pm 0.015$  & $0.16$& $0.165$ \\
 \hline
 $<\bar{\Lambda}+\bar{\Sigma^{0}}>$  & $0.013 \pm 0.01$  & $0.03$ &
  $0.037$  \\
 \hline
 $<K_{s}^{0}>$  & $0.17 \pm 0.01$ & $ 0.26$& $0.27$  \\
 \hline
 $<p>$  & $ 1.34\pm 0.15 $        &$1.43$ & $1.45$ \\
 \hline
 $<\bar{p}>$  & $0.05 \pm 0.02$ & $0.11$& $0.12$  \\
 \hline
 \hline
 \end{tabular}
 \end{table}

\begin{table}
\caption{Particle composition of $p+\bar{p}$
  interactions at 540 GeV in cm.}
\label{Tab2}
\begin{tabular}{||c||c|c|c||}  \hline\hline
{\bf Particle type} & ${\bf <n>}$  & $ {\bf Exp.data}$
& $ {\bf HIJING^{(j)}}$  \\
\hline
\hline
{\bf All charged} &  $29.4 \pm 0.3$ &\cite{1ua5} &  $28.2$  \\
\hline
$K^{0}+\bar{K^{0}}$ & $2.24 \pm 0.16$ & \cite{1ua5} & $1.98$ \\
\hline
$K^{+}+K^{-}$&$ 2.24 \pm 0.16 $ & \cite{1ua5} & $2.06$ \\
\hline
$p+\bar{p}$ & $1.45 \pm 0.15$  & \cite{ward} &  $1.55$ \\
\hline
$\Lambda+\bar{\Lambda}$ & $0.53 \pm 0.11$  & \cite{1ua5} & $0.50$ \\
\hline$\Sigma^{+}+\Sigma^{-}+\bar{\Sigma^{+}}+\bar{\Sigma^{-}}$  &
 $0.27 \pm 0.06$  &    \cite{ward}    & $0.23$  \\
 \hline
 $\Xi^{-}$  &  $0.04 \pm 0.01$ & \cite{1ua5} & $0.037$  \\
 \hline
 $\gamma$  & $33 \pm 3$  & \cite{1ua5} &  $29.02$  \\
 \hline
 $\pi^{+}+\pi^{-}$  &  $23.9 \pm 0.4$ &\cite{1ua5} &   $23.29$ \\
 \hline
 $K_{s}^{0}$ &  $1.1 \pm 0.1$& \cite{1ua5}  &   $0.99$  \\
 \hline
 $\pi^{0}$ & $11.0\pm 0.4$  & \cite{ward} &    $13.36$ \\
 \hline
 \hline
 \end{tabular}
 \end{table}

 \begin{table}
 \caption{Average multiplicities for negative charged hadrons and
 neutral strange hadrons in $\,pp\,$ and $\,pA\,$
 interactions.  HIJING and VENUS  model results are compared
 with others recent estimates using  RQMD, QGSM, DPM
  and with
 data from Alber et al.\protect{\cite{na3594}}.}
\label{Tab3}
\begin{tabular}{||c||c|c|c|c|c||}  \hline \hline
    &   &  &   &   &    \\
 ${ \bf  Reaction}$ &     &${ \bf  <h^{-}>}$ &
 $ { \bf  <\Lambda>}$ & ${ \bf <\bar{\Lambda}>}$
  &${ \bf  <K_{s}^{0}> }$ \\
 \hline
 \hline
       & {\bf DATA} & $ 2.85 \pm 0.03$ &$ 0.096 \pm 0.015$ &
      $ 0.013 \pm 0.005$ & $0.17 \pm 0.01$  \\
      &{\bf HIJING} &$ 2.99$ &$ 0.16$ &$ 0.030$ &$ 0.26  $  \\
      &{\bf VENUS}  & $2.79$   & $0.181$& $0.033$ &$0.27$  \\
     & {\bf RQMD} &$2.59$ & $0.11$ &    &$ 0.21$   \\
 ${\bf p+p}$  & {\bf QGSM} & $2.85$ & $0.15$ & $0.015$ & $0.21$  \\
     &$ {\bf DPM^{1}}$ & $3.52$ & $0.155$ & $0.024$ & $0.18$ \\
     &${\bf DPM^{2}}$ & $3.52$ & $0.155$ & $0.024$ & $0.18$ \\
 \hline
 \hline
 ${\bf  p+S}$ &{\bf DATA} &$ 5.7 \pm 0.2$ &$ 0.28 \pm 0.03$ &
 $ 0.049 \pm 0.006$ & $0.38\pm0.05$ \\
 $'min. bias'  $ & ${\bf HIJING}$ &$4.83$  & $0.255$ & $0.046$ &
 $0.400 $ \\
      & {\bf VENUS}& $5.40$   & $0.340$ &$0.065$ & $0.510$ \\
      & {\bf QGSM} & $5.87$ & $0.240$ & $0.023$ & $0.340$ \\
      &$ {\bf DPM^{1}}$ & $5.53$ & $0.300$ & $0.043$ &$ 0.360$ \\
      &${\bf DPM^{2}}$ & $5.54$ & $0.32$ & $0.060$ &$ 0.360$ \\
 \hline
 \hline
${\bf  p+Ag}$& {\bf DATA} & $6.2\pm0.2$ & $0.37\pm0.06$ &
 $ 0.05 \pm 0.02 $ & $0.525 \pm 0.07 $ \\
 $'min. bias' $ & ${\bf HIJING}$ & $6.28$ & $0.34$ & $0.054$ &$ 0.505$\\
 \hline
 \hline
 ${\bf  p+Au}$& {\bf DATA} & $9.6 \pm 0.2$ &    &    &     \\
 $'central'$ & ${\bf HIJING}$ & $11.25$ & $0.67$ &$ 0.090$ & $0.88$  \\
\hline
\hline
 \end{tabular}
 \end{table}

\begin{table}
 \caption{Average multiplicities for negative charged hadrons and
 neutral strange hadrons in $\,AA\,$
 interactions.  HIJING and VENUS  model results are compared
 with others recent estimates using  RQMD, QGSM, DPM
  and with
 data from Alber et al.\protect{\cite{na3594}}.}

\label{Tab4}
\begin{tabular}{||c||c|c|c|c|c||}  \hline \hline
    &   &  &   &   &    \\
 ${ \bf  Reaction}$ &     &${ \bf  <h^{-}>}$ &
 $ { \bf  <\Lambda>}$ & ${ \bf <\bar{\Lambda}>}$
  &${ \bf  <K_{s}^{0}> }$ \\
 \hline
 \hline
       & {\bf DATA} & $95 \pm 5 $ & $9.4\pm1.0$ &$ 2.2\pm0.4$ &
  $10.5\pm1.7 $\\
 ${\bf  S+S}$   &${\bf HIJING}$ &$88.8$ &$4.58$ & $0.86$ &
 $ 7.23$  \\
  $'central'$ & ${\bf VENUS}$  & $ 94.06$ & $8.20$ & $2.26$ & $11.94$ \\
        &${\bf RQMD}$ & $110.2$ & $7.76$ &   & $10.0$  \\
        &${\bf QGSM}$ & $120.0$ & $4.70$ & $0.35$ & $7.0$ \\
        &${\bf DPM^{1}}$ & $109.8$ & $6.83$ & $0.80$& $10.6$ \\
        &${\bf DPM^{2}}$ & $107.0$ & $7.18$ & $1.57$ & $10.24$ \\
\hline
\hline
      & {\bf DATA} & $160\pm8$ & $15.2 \pm 1.2 $ &
        $ 2.6 \pm 0.3$ &$ 15.5 \pm 1.5 $ \\
${\bf  S+Ag}$ & ${\bf HIJING}$& $ 164.35$ &$ 8.61$ &
   $1.48$ & $13.20$ \\
$'central'$ & ${\bf RQMD}$ & $192.3$ & $13.4$ &    &  $18.30$ \\
   &${\bf DPM^{1}}$ & $195.0$ & $13.3$ & $1.45$ & $19.40$ \\
   &${\bf DPM^{2}}$ & $186.90$ & $14.06$ & $3.65$ & $15.73$ \\
\hline
\hline
${\bf  S+Au}$ & ${\bf HIJING}$ & $213.2$ & $11.3$ & $1.81$ & $16.55$ \\
$'central'$ &  ${\bf VENUS}$  & $201.6$ & $14.0$ & $3.01$ & $21.52$ \\
 \hline
 \hline
${\bf  S+W}$ & ${\bf HIJING}$ & $210.0$ & $10.64$ &$ 1.71$ &$16.05$ \\
$'central'$&   &    &    &     &   \\
 \hline
 \hline
${\bf  Pb+Pb}$  & ${\bf HIJING}$ &$725.15$ & $36.44$ &$5.93$ & $54.86$ \\
$'central'$&    &    &      &    &      \\
\hline
\hline
 \end{tabular}
 \end{table}

\begin{table}
\caption{The mean multiplicities of negative pions and $\,\,E_{S}\,\,$
 ratios(see the text for definition) for nuclear collisions
 at $\,\, 200\,$ AGeV. The data are from
 Alber et al.\protect{\cite{na3594}}and the $\,\,NN\,\,$ data are from
 Gazdzicki and Hansen \protect{\cite{11na35}}.}
\label{Tab5}
 \begin{tabular}{||c||c|c|c||}  \hline \hline
 ${\bf Reaction}$ &     &  $<\pi^{-}>$ &
 $<E_{S}>$ \\
 \hline
 \hline
  ${\bf p+p}$  & {\bf DATA} & $ 2.62 \pm 0.06$ &    \\
      &{\bf HIJING} & $ 2.61$ & $0.153$   \\
 \hline
 \hline
 ${\bf N+N}$ &{\bf DATA}&  $3.06 \pm 0.08 $ & $0.100\pm 0.01$   \\
        & $ {\bf HIJING}$ &$2.89$  & $0.140$       \\
 \hline
 \hline
 ${\bf  p+S}$ &{\bf DATA} &$5.26\pm0.13$ &$ 0.086 \pm 0.008$   \\
 $'min. bias'  $ & ${\bf HIJING}$ &$4.3$  & $0.144$     \\
 \hline
 \hline
${\bf  p+Ag}$& {\bf DATA} & $6.4\pm0.11$ & $0.108\pm0.009$   \\
 $'min. bias' $ & ${\bf HIJING}$ & $5.59$ & $0.141$      \\
 \hline
 \hline
 ${\bf  p+Au}$& {\bf DATA} & $9.3 \pm 0.2$ &$0.073\pm0.015$     \\
 $'central'$ & ${\bf HIJING}$ & $10.22$ & $0.136$   \\
 \hline
 \hline
     ${\bf  S+S}$  & {\bf DATA} & $88\pm5 $ & $0.183\pm0.012$    \\
   $'central'$   & ${\bf HIJING}$ & $79.6$ & $0.140$   \\
\hline
\hline
 ${\bf  S+Ag}$ & {\bf DATA} & $149\pm8$ & $0.173\pm0.017$  \\
  $'central'$  & ${\bf HIJING}$ &$147.8$   &$0.138$   \\
\hline
\hline
 \end{tabular}
 \end{table}


\newpage

\begin{thebibliography}{199}
\bibitem{pro6}{\it Proceedings of the Tenth International
Conference on Ultra-Relativistic Nucleus-Nucleus Collisions
-Quark Matter '93,Borlange,Sweden ,June 20--24,1993}
Edited by E.~Stenlund,H.~A.~Gustafsson,A.~Oskarsson and
I.~Otterlund\\
Nucl.Phys.{\bf A566}(1994)
\bibitem{pro5}{\it Proceedings of a NATO Advanced
Study Institute on Particle Production in Highly Excited Matter,
July 12--24,1992 IL CIOCCO,Tuscany,Italy\/}Eds .Hans.H.Gutbrod
and Johann Rafelski,Plenum Press New York and London (1993)
\bibitem{raf1}P.Koch.B.Muller and J.Rafelski
Phys.Rep.{\bf C142}167(1986)
\bibitem{quer1}E.Quercigh,{\it Proceedings of a NATO Advanced
Study Institute on Particle Production in Highly Excited Matter,
July 12--24,1992 IL CIOCCO,Tuscany,Italy\/}Eds .Hans.H.Gutbrod
and Johann Rafelski,Plenum Press New York and London (1993),p499
\bibitem{rafel} J. Letessier, J. Rafelski, Ahmed Tounsi,
 Phys.Lett.{\bf B323},393(1994);Preprint PAR-LPTHE-92-27-REV unpublished.
\bibitem{quer2}E.~Quercigh,
Nucl.~Phys.{\bf A566},321c(1994)
\bibitem{1na35}J.Bartke et al.,Z.Phys.{\bf C48},191(1990)
\bibitem{2na35}R.Stock et al.,Nucl.Phys.{\bf
A525},221c(1991)
\bibitem{3na35}P.Seyboth et al.,Nucl.Phys.{\bf
A544},293c(1992)
\bibitem{4na35}M.Kowalski et at.,Nucl.Phys.{\bf
A544},609c(1992)
\bibitem{5na35}H.Bialkowska et al.,Z.Phys.{\bf
C55},491(1992)
\bibitem{6na35}J.Baechler et al.,Z.Phys.{\bf
C58},367(1993)
\bibitem{9na35}A.Bamberger et al.,Z.Phys.{\bf
C43},25(1989)
\bibitem{10na35}I.Derado et al.,Z.Phys.{\bf C50},
31(1991)
\bibitem{11na35}M.~Gazdzicki and O.~Hansen,Nucl.~Phys.,
{\bf A528},754(1991)
\bibitem{gaz94} M.~Gazdzicki et al.,
Nucl.~Phys.~{\bf A566},503c(1994)
\bibitem{12na35} H.Strobele et al.,
Nucl.Phys.{\bf A525},59c(1991)
\bibitem{na3594} T.Alber et.al,
Z.fur Physik {\bf C64},195(1994)
\bibitem{1ander94} E.~Andersen et.~al.~,
Nucl.~Phys.~{\bf A566},217c(1994)
\bibitem{2ander94} E.Andersen et.~al.~,
Nucl.~Phys.~{\bf A566},487c(1994)
\bibitem{1na36}D.E.Greiner et al.,Nucl.Phys.{\bf
A544},309c(1992)
\bibitem{2na36}E.Anderson et al.,Nucl.Phys.{\bf
A553},817c(1993)
\bibitem{3na36}E.Andersen et al.,Phys.Lett.{\bf
B294},127(1992)
\bibitem{5na36}E.Andersen et al.,Phys.Lett.{\bf
B316},603(1993)
\bibitem{6na36}E.Andersen et al.,Phys.Lett.{\bf
B327},433(1944)
\bibitem{greiner95}D. Greiner, et al, (NA36) preprint  LBL-36882 (1995)
unpublished.
\bibitem{na36qm95} E.Judd (NA36 Collaboration),Proceedings of the
Eleventh International Conference on Ultra-Relativistic
Nucleus-Nucleus Collisions,January 9-13,1995,Monterey,California,
USA, Nucl.Phys.{\bf A} (in press)
\bibitem{1abat94} S.Abatzis et.~al.
 Nucl.~Phys.~{\bf A566},225c(1994)
\bibitem{2abat94} F.~Antinori et.~al.~,
Nucl.~Phys.~{\bf A566},491c(1994)
\bibitem{3abat94} S.Abatzis et.~al.~,
 Nucl.~Phys.~{\bf A566},499c(1994)
\bibitem{1wa85}S.Abatzis et al.,Phys.Lett.{\bf
B244},130(1990)
\bibitem{2wa85}S.Abatzis et al.,Phys.Lett.{\bf
B259},508(1991)
\bibitem{3wa85}S.Abatzis et al.,Phys.Lett.{\bf
B270},123(1991)
\bibitem{4wa85}S.Abatzis et al.,Nucl.Phys.{\bf
A525},445c(1991)
\bibitem{6wa85}J.B.Kinson et al.,Nucl.Phys.{\bf
A544},321c(1992)
\bibitem{5hel}T.Akesson et al.,Z.~Phys.{\bf C53},183(1992)
\bibitem{sar92}M.~Sarabura et al.,Nucl.~Phys.
{}~{\bf A544},125c(1992)
\bibitem{mur94}M.~Murray et al.,Nucl.~Phys.
{}~{\bf A566},515c(1994)
\bibitem{3e802}T.Abbot et al.,Phys.Rev.{\bf D45},3906(1992)
\bibitem{4e802}Y.Miake et al.,Nucl.Phys.{\bf A525},231c(1991)
\bibitem{5e802}T.Abbot et al.,Phys.Rev.{\bf C47},1351(1993)
\bibitem{1e810}K.J.Foley et al.,Nucl.Phys.{\bf A544},325c(1992)
\bibitem{e810long} R.~S.~Longacre et al.,
Nucl.~Phys.{\bf A566},167c(1994)
\bibitem{e814stac} Johana Stachel et.~al.~,
Nucl.Phys.{\bf A566},183(1994)
\bibitem{bare94}J.~Barette et al.,Z.Phys.{\bf C59},211(1993)
\bibitem{dieb93} G.~E.~Diebold et.~al.~,
Phys.~Rev.~{\bf C48},2984(1993)
\bibitem{nam94} M.~N.~Namboodiri et.~al.~,
Nucl.~Phys.{\bf A566},443c(1994)
\bibitem{dpm94} A.Capella,U.Sukhatme,C.~I.~Tan and
J.~Tran ~Thanh ~Van ,Phys.~Rep.{\bf 236},225(1994)
\bibitem{mohr1}H.~J.~Mohring,J.Ranft,A.Capella and J.Tran Thanh Van,\\
Phys.Rev.{\bf D47},4146(1993)
\bibitem{mohr2}H.~J.~Mohring J.Ranft,C.Merino and C.Pajares,Phys.Rev.
{\bf D47},4142(1993)
\bibitem{ranft1}J.Ranft and S.Ritter,Z.Phys.{\bf C27},
413(1985)
\bibitem{ranft2}H.J.Mohring,J.Ranft,S.Ritter,
Z.Phys.{\bf C27},419(1985)
\bibitem{ranft3}H.J.Mohring,J.Ranft,Z.Phys.{\bf C52},
643(1991)
\bibitem{ranft4}A.Capella,C.Merino,H.J.Mohring,J.Ranft,J.Tran Thanh Van,\\
Nucl.Phys.{\bf A525},493c(1991)
\bibitem{kai1}A.B.Kaidalov,Sov.J.Nucl.Phys.{\bf
33},733(1981)
\bibitem{kai2}A.B.Kaidalov and O.I.Piskunova,Z.Phys.{\bf
C30},145(1986)
\bibitem{kai3}A.B.Kaidalov ,Nucl.Phys.{\bf A525},39c(1991)
\bibitem{ame1}N.~S.~Amelin ,E.F.Staubo,L.P.Csernai,V.D.Toneev,K.K.Gudima, \\
Phys.Rev.{\bf C44},1541(1991)
\bibitem{ame2}N.~S.~Amelin,L.V.Bravina,L.P.Csernai,V.D.Toneev
K.K.Gudima,S.Yu. Sivoklokov,\\
Phys.Rev.{\bf C47},2299(1993)
\bibitem{wer7}K.~Werner,Preprint HD-TVP-93-1(1993),University of
Heidelberg,Germany;\\
Phys.Rep.{\bf 232},87(1993)
\bibitem{ander}B.Andersson,G.Gustafson and
Nilsson-Almqvist Nucl.Phys.{\bf B281},289(1987)
\bibitem{nils}B.Nilsson-Almqvistand E.Stenlund,
Comp.Phys.Commun.{\bf 43},387(1987)
\bibitem{Gyu1}M.~Gyulassy ,CERN Report TH-4794/87(1987),in Proc.of Int.
 Conf. on Intermediate Energy Nucl.Phys,Balatonfured,Hungary,1987
\bibitem{wang0}Xin-Nian Wang and Miklos Gyulassy,
Comp.Phys.Comm.{\bf 83},307(1994)
\bibitem{wang1}Xin-Nian Wang,Phys.Rev.{\bf D43},104(1991)
\bibitem{wang2}Xin-Nian Wang and Miklos Gyulassy,Phys.Rev.
{\bf D44},3501(1991)
\bibitem{wang3}Xin-Nian Wang and Miklos Gyulassy,Phys.Rev.
{\bf D45},844(1992)
\bibitem{wang4}Xin-Nian Wang and Miklos Gyulassy,Phys.Rev.Lett.
{\bf 68},1480(1992)
\bibitem{aich1}J.~Aichelin,G.Peilert,A.Bohnet,A.Rosenhauer,H.Socker
and W.Greiner ,\\
Phys.Rev.{\bf C37},2451(1988)
\bibitem{sorge1}H.~Sorge,H.~Stocker and W.~Greiner,
Nucl.Phys.{\bf A498},567c(1989)
\bibitem{sorge3} H.Sorge,L.A.Winckelmann,H.Stocker,W.Greiner,
Z.fur Phys.{\bf C49},85(1993)
\bibitem{sor93a}H.~Sorge ,Preprint {\bf
LA-UR-93-1103}(1993)Los Alamos,USA
\bibitem{sor94a}H.~Sorge,Nucl.Phys.{\bf A566},633c(1994)
\bibitem{sor94}H.~Sorge ,Phys.Rev.{\bf C49},1253(1994)
\bibitem{ame94}N.S.Amelin,H.Stocker,W.Greiner,
N.Armesto,M.A.Braun and C.Pajares,
Preprint {\bf US-FT/1-94}(Santiago University)
(1994)(subm.to Phys.Rev.{\bf C})
\bibitem{fol94} R.~Folman,A.~Shor ,
Nucl.~Phys.{\bf A566},917(1994)
\bibitem{gei1} K.Geiger and B.Mueller,Nucl.Phys.{\bf B369},600(1992)
\bibitem{gei2} K.Geiger, Phys.Rev.{\bf D47},133(1993)
\bibitem{gei3}K.Geiger,Preprint CERN, CERN - TH -7313(1994)
(subm.to Phys.Rep.)
\bibitem{braun4}M.~Braun and C.~Pajares,~Phys.~Rev.{\bf D47},114(1993)
\bibitem{merino1}C.~Merino,C.Pajares and
J.Ranft,Phys.Lett.{\bf B276},168(1992)
\bibitem{merino2}H.~Mohring,J.Ranft,C.Merino and
C.Pajares ,Phys.Rev.{\bf D47},4142(1993)
\bibitem{biro} T.S. Biro, H.B. Nielsen, J. Knoll,
Nucl. Phys. {\bf B245},449(1984)
\bibitem{sjos94}T.Sjostrand,Comp.Phys.Comm.{\bf 82},74(1994)
\bibitem{hwa1}R.~C.~Hwa,Phys.Rev.{\bf D37},1830(1988)
\bibitem{1ua5} G.J.Alner et al.,Phys.Rep.{\bf 154},247(1987)
\bibitem{ward} D.~R.~Ward {\it Properties of Soft
Proton Antiproton Collisions }in Advances Series on
Direction in High Energy Physics ,Vol.4 ,Eds.G.Altarelli and
L.Di Lella,1989 ,p.85
\bibitem{drem} I.~M.~Dremin ,Sov.J.Nucl.Phys.
{\bf 33},726(1981)
\bibitem{jagl} Jaeger et al.,Phys.Rev.{\bf D11},2405(1975)

\end{thebibliography}
 
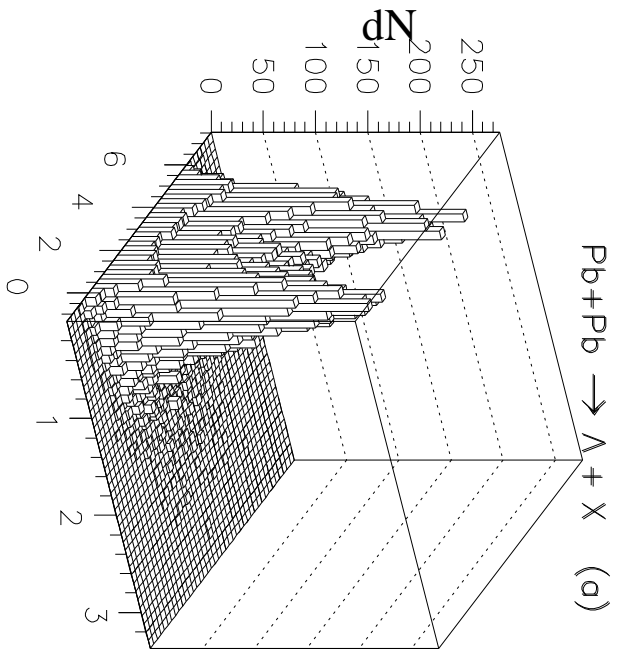
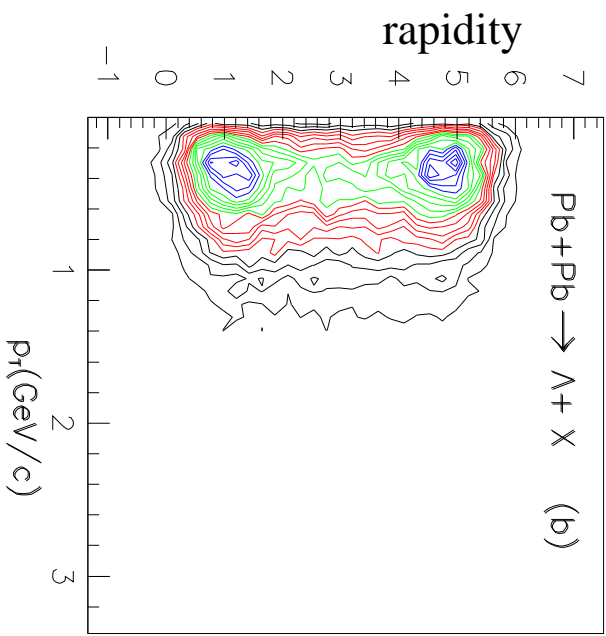
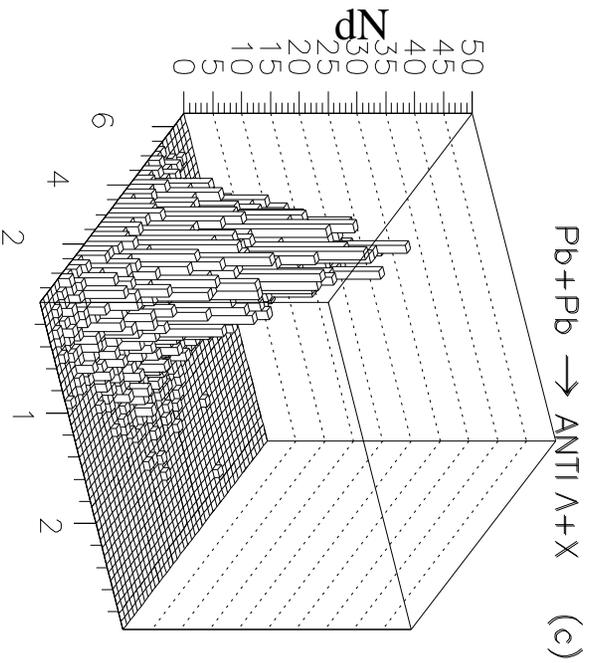
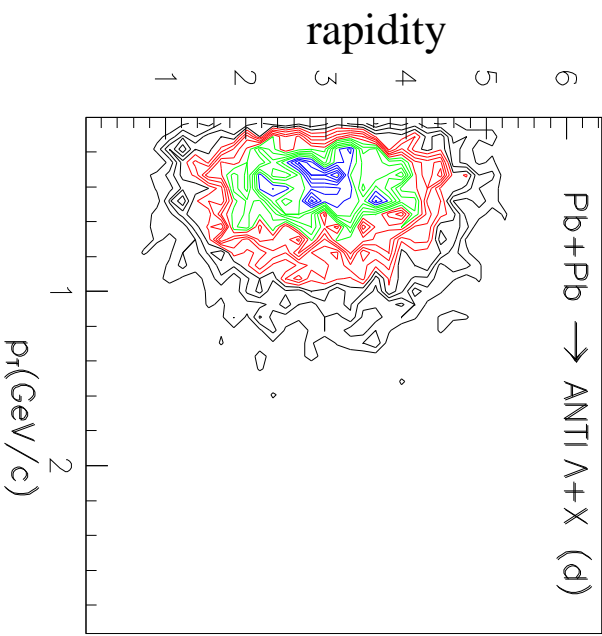
\end{document}